\documentclass[pra,twocolumn,superscriptaddress]{revtex4-2}
\usepackage{amsmath}
\usepackage{color}
\usepackage{epsfig,graphicx,amssymb,bm}
\usepackage{subfigure}

\begin{document}

\title{Synchronization by Magnetostriction}
\author{Jiong Cheng}
\affiliation{Interdisciplinary Center of Quantum Information, State Key Laboratory of Modern Optical Instrumentation, and Zhejiang Province Key Laboratory of Quantum Technology and Device, School of Physics, Zhejiang University, Hangzhou 310027, China}
\affiliation{Department of Physics, School of Physical Science and Technology, Ningbo University, Ningbo, 315211, China}
\author{Wenlin Li}\thanks{liwenlin@mail.neu.edu.cn}
\affiliation{College of Sciences, Northeastern University, Shenyang 110819, China}
\author{Jie Li}\thanks{jieli007@zju.edu.cn}
\affiliation{Interdisciplinary Center of Quantum Information, State Key Laboratory of Modern Optical Instrumentation, and Zhejiang Province Key Laboratory of Quantum Technology and Device, School of Physics, Zhejiang University, Hangzhou 310027, China}
%\date{\today}

\begin{abstract}
We show how to utilize magnetostriction to synchronize two mechanical vibration modes in a cavity magnomechanical system. The dispersive magnetostrictive interaction provides necessary nonlinearity required for achieving synchronization. Strong phase correlation between two mechanical oscillators can be established, leading to the synchronization robust against thermal noise. We develop a theoretical framework to analyze the synchronization by solving the constraint conditions of steady-state limit cycles. We determine that the strong cavity-magnon linear coupling can enhance and regulate the synchronization, which offers a new path to modulate synchronization. The work reveals a new mechanism for achieving and modulating synchronization and indicates that cavity magnomechanical systems can be an ideal platform to explore rich synchronization phenomena.
\end{abstract}

%\pacs{03.65.Yz,42.50.Wk, 03.65.Ud}

\maketitle

\section{Introduction}

The emergence of spontaneous order in coupled systems, known as spontaneous synchronization, is a ubiquitous phenomenon in various natural and social systems~\cite{Pikovsky2001,Osipov2007}. Over the past few decades, synchronization phenomena have been thoroughly investigated in the classical domain~\cite{Acebron137,Arenas93}.  In recent years, researches in this field have been gradually extended into the microcosmic regime~\cite{Lee234101,Witthaut14829,Cabot2019,Sonar163601,Schmolke2022,Laskar2020,Vinokur2008}, where quantum effects, e.g., quantum fluctuations and the Heisenberg uncertainty principle~\cite{Mari103605,Li022204}, nonclassical properties of the non-Gaussian states~\cite{Lee234101,Sonar163601,Li023512}, quantum correlations~\cite{Giorgi052101,Roulet2018,Solanki2022,Ameri012301}, quantum phase transitions~\cite{Pizzi094301,Schmolke2022}, etc, manifest themselves.  Subsequently, the phenomena have been systematically explored and summarized as the quantum synchronization theory, which also reveals the deep mechanisms of some remarkable quantum effects~\cite{Schmolke2022,Ludwig073603,Richerme2017} and provides a new perspective on fundamental quantum theories~\cite{Ludwig073603,Li023512} and quantum information processing~\cite{Roulet2018,Li022204,Nande109772}. Synchronization in various microcosmic systems have been observed or predicted, e.g., in subatomic particle ensembles~\cite{Laskar2020,Xu2014,Schmolke2022}, mechanical resonators~\cite{Zhang233906,Ludwig073603,Bagheri213902,Mari103605,Zhang163902,Santos063605,Li022204,Sonar163601,Colombano2019,Li013802,Sheng2020,Piergentili073013,Li023512},
and cavity or circuit electrodynamics systems~\cite{Ameri012301,Roulet2018}. All of them correspond to complex models with multiple subsystems, or eigenmodes, coupled by appreciable nonlinear interactions (strong enough, typically enhanced by an intense pump, to support self-sustaining dynamics~\cite{Roulet053601}). Among them, only a few systems can be well analyzed beyond the purely numerical results, and unfortunately, the constraints imposed by current experimental techniques further narrow the range of such candidate systems~\cite{Laskar2020,Zhang233906,Santos063605,Bagheri213902,Zhang163902,Sheng2020,Colombano2019,Piergentili073013}. A mature and easy-to-control platform capable of bridging synchronization theory, numerical analysis and  experimental observation is highly desired.

Here, we show that the recently developed cavity magnomechanical (CMM) system~\cite{Tang,Jie18,Davis,Jie22} can exactly be such a candidate system. In the CMM system, magnons, quanta of collective spin excitations, in a ferrimagnetic yttrium-iron-garnet (YIG) sphere couple to vibration phonons via the magnetostrictive interaction, which is a dispersive interaction~\cite{Kittel836,QST} and thus provides necessary nonlinearity for achieving synchronization in the system. Such nonlinearity also plays an essential role in preparing macroscopic quantum states~\cite{Jie18,Li021801,Li085001,Tan,Li024005} and designing novel quantum technologies~\cite{Naka19,Yuan965,Kong5544,Jie20,Potts064001,Jing,Sarma043041,Qi043704,NSR}.  In addition, magnons further couple to microwave cavity photons via the magnetic-dipole interaction. Due to the high spin density of YIG, the strong cavity-magnon coupling can be easily achieved, leading to cavity polaritons~\cite{Huebl127003,Tabuchi083603,Zhang156401}. Such a coupling is adjustable by changing the position of the YIG sphere in the microwave cavity. The intrinsic nonlinearity and tunable strong coupling of the CMM system make it an ideal platform to explore synchronization.

Specifically, we show that it is possible to achieve robust synchronization of two mechanical vibration modes protected by strong phase correlation under feasible parameters even at room temperature. The synchronization in the CMM system can be analytically decomposed by mapping the constraint conditions of steady-state limit cycles into the parameter space, which provides us a simple way to understand the complicated dynamics of the synchronization.  We find that the strong cavity-magnon coupling provides a new degree of freedom, which plays an important and active role in enhancing and modulating the synchronization. This represents a new path to the modulation of synchronization and fundamentally differs from the synchronization mechanism in other systems, e.g., optomechanical systems~\cite{Zhang233906,Bagheri213902,Zhang163902,Santos063605,Colombano2019,Li013802,Sheng2020,Piergentili073013}.

\begin{figure}[t]
\includegraphics[width=8.25cm]{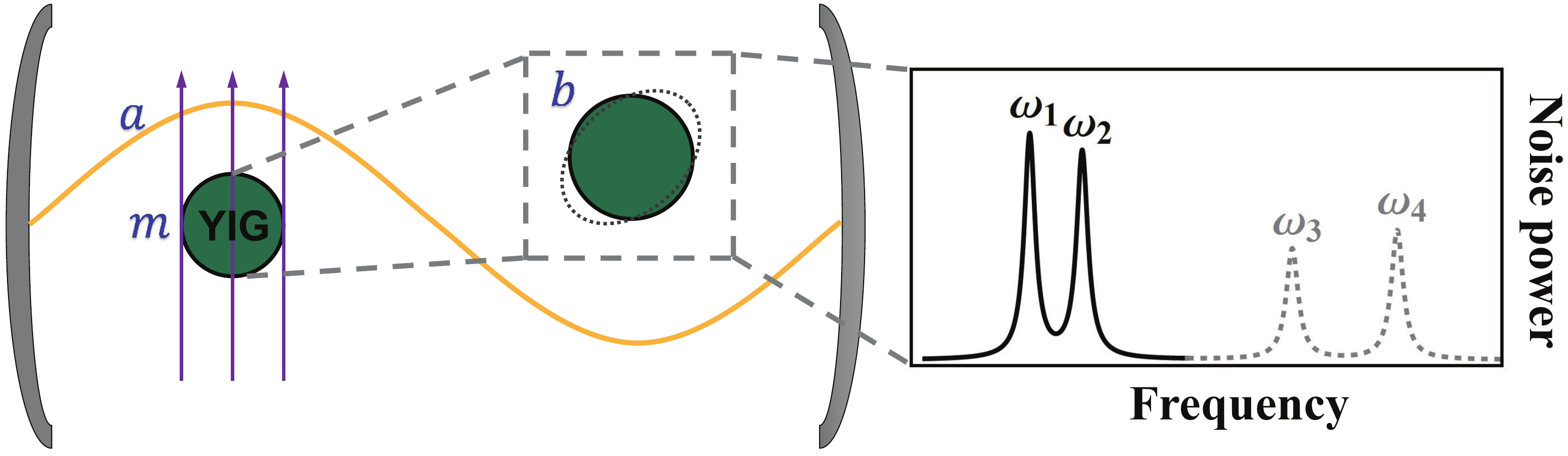}
\caption{Schematic diagram of the cavity magnomechanical system used for achieving synchronization of two mechanical modes. It consists of a microwave cavity mode $a$, a magnon mode $m$, and two mechanical vibration modes $b_{1,2}$ (with resonance frequencies of $\omega_{1,2}$). }
\label{model}
\end{figure}

\section{The model}

We consider a typical cavity magnomechanical system~\cite{Tang,Jie18,Davis,Jie22}, as depicted in Fig.~\ref{model}. It consists of a microwave cavity and a macroscopic YIG sphere placed inside the cavity, which supports a magnon (spin wave) mode and a series of mechanical vibration modes, among which we focus on two mechanical modes and study the synchronization between them.  %The collective excitations of a large number of spins inside the YIG sphere is quantized as the magnon mode, while the mechanical mode is the deformation mode of the sphere caused by the magnetostrictive force~\cite{Kittel836,Tang,Jie18,Li021801}, which also lead to a nonlinear
%radiation pressure-like magnomechanical interaction between magnons and phonons when the mechanical frequency is much smaller than the magnon frequency~\cite{Tang}.Meanwhile, the magnetic dipole interaction mediates the coupling between the magnon mode and the microwave cavity mode.
The Hamiltonian of the system reads:
\begin{eqnarray}
\hat{H}/\hbar &=& \omega_{a}\hat{a}^{\dag}\hat{a}+\omega_{m}\hat{m}^{\dag}\hat{m}
+g_{ma}(\hat{a}^{\dag}\hat{m}+\hat{m}^{\dag}\hat{a}) \notag \\
&&+\sum_{j=1,2} \left[ \omega_{j}\hat{b}_{j}^{\dag}\hat{b}_{j}
+g_{j}\hat{m}^{\dag}\hat{m}(\hat{b}_{j}^{\dag}+\hat{b}_{j}) \right] \notag \\
&&+i \Omega (\hat{m}^{\dag}e^{-i\omega_{0}t}-\hat{m}e^{i\omega_{0}t}),
\label{Hamiltonian}
\end{eqnarray}where $\hat{a}$, $\hat{m}$ and $\hat{b}_{j}$ ($\omega_{a}$, $\omega_{m}$, and $\omega_{j}$) are the annihilation operators (resonance frequencies) of the cavity, magnon and 
$j$-th mechanical modes, respectively, satisfying $[\hat{O},\hat{O}^{\dag }] \,{=}\, 1$ $(\hat{O}=\hat{a},\hat{m},\hat{b}_{j})$.
The magnon frequency can be adjusted by altering the bias magnetic field $H_0$ via $\omega_{m}=\gamma_{0}H_0$, with the gyromagnetic ratio $\gamma_{0}/2\pi=28$ GHz/T.  $g_j$ denotes the bare coupling rate between the magnon and the $j$-th mechanical mode and $g_{ma}$ is the cavity-magnon coupling rate, which can be (much) stronger than the cavity and magnon dissipation rates $\kappa_{a}$ and $\kappa_{m}$~\cite{Huebl127003,Tabuchi083603,Zhang156401}.  To enhance the magnetostrictive interaction, the magnon mode is driven by a microwave field with frequency $\omega_{0}$ and amplitude $B_{0}$, and the corresponding Rabi frequency is $\Omega=(\sqrt{5}/4)\gamma_{0}\sqrt{N}B_{0}$~\cite{Jie18}, where $N=\rho V$ is the total number of spins, $\rho=4.22\times10^{27}$~${\rm m}^{-3}$ is the spin density of YIG, and $V$ is the volume of the sphere.

In the frame rotating at the driving frequency $\omega_{0}$, and by adding dissipative and input noise terms, we obtain the following quantum Langevin equations (QLEs):
\begin{eqnarray}
\dot{\hat{a}} &=& -(i\Delta_{a}+\kappa_{a})\hat{a}-ig_{ma}\hat{m}+\sqrt{2\kappa_{a}}\hat{a}^{in}, \notag \\
\dot{\hat{m}} &=& -(i\Delta_{m}+\kappa_{m})\hat{m}-ig_{ma}\hat{a}-\sum_{j=1,2}ig_{j}\hat{m}(\hat{b}_{j}^{\dag}+\hat{b}_{j}) \notag \\
&&+\Omega+\sqrt{2\kappa_{m}}\hat{m}^{in},\notag \\
\dot{\hat{b}}_{j} &=& -(i\omega_{j}+\gamma_{j})\hat{b}_{j}-ig_{j}\hat{m}^{\dag}\hat{m}+\sqrt{2\gamma_{j}}\hat{b}^{in}_{j},
\label{QLEs}
\end{eqnarray}
where $\Delta_{a}=\omega_{a}-\omega_{0}$ and $\Delta_{m}=\omega_{m}-\omega_{0}$.
$\kappa_{a}$, $\kappa_{m}$ and $\gamma_{j}$ ($\hat{a}^{in}$, $\hat{m}^{in}$ and $\hat{b}^{in}_{j}$) are the decay rates (input noise operators) of the cavity, magnon and $j$-th mechanical modes, respectively.
The input noises are assumed Gaussian and white noises, of which the correlation functions are
$\langle\hat{O}^{in}(t)\hat{O}^{in\dag}(t^{\prime})\rangle=(\bar{N}_{O}+1)\delta(t-t^{\prime})$ and $\langle\hat{O}^{in\dag}(t)\hat{O}^{in}(t^{\prime})\rangle=\bar{N}_{O}\delta(t-t^{\prime})$, with $\hat{O}=\hat{a}$, $\hat{m}$, $\hat{b}_{j}$, and $\bar{N}_{O}=[\exp(\hbar\omega_{O}/k_{B}T)-1]^{-1}$ ($O=a,m,j$) being the mean thermal excitation number of the corresponding mode, $k_{B}$ the Boltzmann constant and $T$ the bath temperature.

%%%%%%%%%%%%%%%%%%%%%%%%%%%%%%%%%%%%%%%%%%%%%%%%%%%%%%%%%%%%%%%%%%%%%%%%%%%%%%%%%%
\section{Phase noise analysis}

To study synchronization at a finite temperature, thermal noises of the system must be included, as the mean thermal occupation $\bar{N}_{O}\gg1$ at a high temperature, e.g., room temperature.  We therefore apply stochastic Langevin equations (operators $\hat{O}$ are replaced with complex variables $O$)~\cite{Weiss013043,Li013802} to describe the system dynamics and simulate them numerically up to the long-time limit. The stochastic Langevin equations associated with Eq. (\ref{QLEs}) are given by~\cite{Wang2014,Benlloch133601}:
\begin{eqnarray}
\dot{a} &=& -(i\Delta_a+\kappa_a)a-ig_{ma}m+\sqrt{2\kappa_a}a^{in}, \notag \\
\dot{m} &=& -i(\Delta_m+\kappa_m)m-ig_{ma}a-\sum_{j=1,2} ig_jm(b_j^*+b_j) \notag \\
&&+\Omega+\sqrt{2\kappa_m}m^{in}, \notag \\
\dot{b}_j &=& -(i\omega_j+\gamma_j)b_j-ig_j\vert m\vert^2+\sqrt{2\gamma_j}b^{in}_j.
\label{eq:cle}
\end{eqnarray}
The operators $\hat{a}$, $\hat{m}$ and $\hat{b}_j$ in the QLEs  are replaced with $c$-number complex variables ${a}$, ${m}$ and ${b_j}$, and the input noise operators are replaced with classical complex random noises with modified correlation functions: $\langle {O}^{in,*}(t){O}^{in}(t')\rangle = (\bar{n}_O+1/2)\delta(t-t')$ ($O\in\{a,m,b_j\}$), because the $c$-numbers lose the commutation relation.

After repeatedly calculating the stochastic Langevin equations $N$ times ($N$ should be large), the disturbance of the noises to the synchronization can be characterized by the phase-space probability distribution of the considered phases and the phase correlation (particularly, the phase difference), which are defined as:
\begin{eqnarray}
P_{\theta_j}(\theta) &=& \lim_{h\rightarrow 0}\dfrac{N_{\theta_j}(\theta)}{N h}, \notag \\
P_{\theta_-}(\theta) &=& \lim_{h\rightarrow 0}\dfrac{N_{\theta_{-}}(\theta)}{N h}, 
\label{eq:modulus distribution function}
\end{eqnarray}
where $N_{\theta_{j(-)}}(\theta)$ is the number of the results satisfying $\theta_{j(-)}^i\in[\theta-h/2,\theta+h/2]$, and the superscript $i$ denotes the $i$-th stochastic trajectory in the simulation, $\theta_{j}$ is the phase of the slowly varying complex amplitude of the $j$-th oscillator (see Appendix A).
The ensemble-averaged quantities and their quantum fluctuations can be estimated by $\langle\theta_{j(-)}\rangle=\sum \theta^i_{j(-)}/N$ and $\langle\theta^2_{j(-)}\rangle=\sum {\theta^{i}_{j(-)}}^2/N-(\sum \theta^i_{j(-)}/N)^2$, respectively.

\begin{figure}[t]
\includegraphics[width=8.5cm]{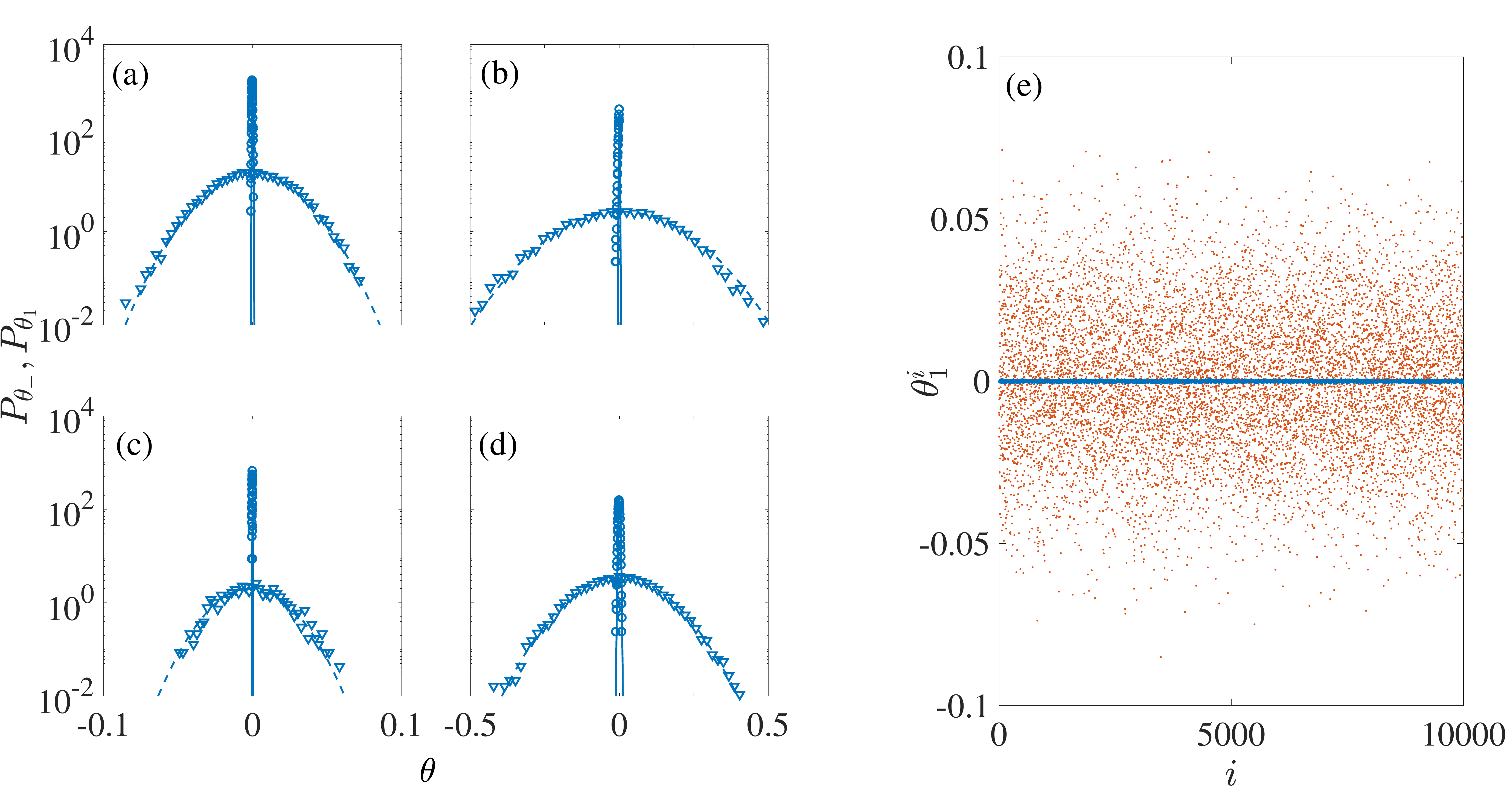}
\caption{(a), (b), (c) and (d): Phase difference probability distribution $\theta_-$ (circle) and phase probability distribution of the oscillator $1$ $\theta_1$ (triangle) obtained by $10^4$ times calculations of the stochastic equations with different values of $\Omega$ and $g_{ma}$ (from left to right and from top to bottom, corresponding to the cases \romannumeral1, \romannumeral2, \romannumeral3~and \romannumeral4, respectively). 
The solid and dashed lines represent the results that fit the corresponding original data with a Gaussian distribution. (e): $10^4$ statistical results of $\theta_-$ (blue) and $\theta_1$ (red) for the case \romannumeral1. 
\label{fig1}}
\end{figure}

%The noise analysis corresponds to the following four cases: \romannumeral1) the synchronization phase; \romannumeral2) the anti-synchronization phase; \romannumeral3) the critical boundary between the synchronization and anti-synchronization; and \romannumeral4) %a weak pump case close to the critical parameters distinguishing the self-sustaining dynamics and asymptotically steady-state dynamics. 
We simulate the stochastic Langevin equations for a time interval $\gamma_1 t=19$, i.e., the same time interval used to obtain the phase diagram in the following section, and repeat the calculations $10^4$ times. The noise analysis corresponds to the following four cases,
and each case corresponds to a specific stable phase difference (or a set of parameters). The state of the system can be well described by the four cases, i.e., the two oscillators are: \romannumeral1) synchronized; \romannumeral2) anti-synchronized; \romannumeral3) at the critical point between synchronization and anti-synchronization; \romannumeral4) anti-synchronized with low energy (see the phase diagram for the detailed description).
The corresponding parameters are respectively: \romannumeral1) $\log_{10}(\Omega/\Omega_0)=-0.4$, $g_{ma}/\omega_1=0.8$;  \romannumeral2) $\log_{10}(\Omega/\Omega_0)=-0.8$, $g_{ma}/\omega_1=0.8$;  \romannumeral3) $\log_{10}(\Omega/\Omega_0)=-0.5168$, $g_{ma}/\omega_1=0.7$ and \romannumeral4) $\log_{10}(\Omega/\Omega_0)=-1$, $g_{ma}/\omega_1=0.5$. We use experimentally feasible parameters in getting Fig.~\ref{fig1}~\cite{Tang,Jie18,Davis,Jie22}: $\omega_{a}=\omega_{m}=2\pi\times10$ GHz, $\omega_{1}=2\pi\times10$ MHz, $\kappa_{a}= 2\pi\times 1.5$ MHz, $\kappa_{m}= 2\pi\times 1$ MHz, $\gamma_{1}=2\pi\times 100$ Hz, $\gamma_{2}=2\pi\times 150$ Hz, $g_{1}=2\pi \times 60$ mHz, $g_{2}=2\pi \times 50$ mHz and $\Omega_{0}=7\times10^{14}$ Hz (corresponding to the drive magnetic field $B_{0}=3.8\times10^{-5}$~T and power $P=8.3$~mW~\cite{Jie18}).

The solid lines in Fig.~\ref{fig1}(a-d) show the phase probability distribution of the mechanical oscillator $1$. We find that when the effects of thermal noises are taken into account, the phase of the  mechanical mode is progressively diffused with phase variance $\langle \delta \theta_1^2\rangle= 4.8\times 10^{-4}$ corresponding to case \romannumeral1, $\langle \delta\theta_1^2\rangle= 2.2\times 10^{-2}$ in case \romannumeral2, $\langle \delta\theta_1^2\rangle= 3.5\times 10^{-4}$ in case \romannumeral3, and $\langle \delta \theta_1^2\rangle= 1.3\times 10^{-2}$ in case \romannumeral4. 
The circles show that the distribution of the phase difference is obviously narrowed, which can be described quantitatively by defining a compression ratio:
\begin{equation}
\begin{split}
\eta=\dfrac{\langle \delta \theta_-^2\rangle}{\langle \delta \theta_1^2\rangle}.
\end{split}
\label{eq:modulus distribution function}
\end{equation}
$\eta$ equals to $1$ for two uncorrelated oscillators due to $\langle \delta \theta_-^2\rangle\simeq \langle \delta \theta_1^2\rangle\simeq \langle \delta \theta_2^2\rangle$ in this case. We obtain $\eta=1.19\times 10^{-4}$, $1.16\times 10^{-4}$,  $1.75\times 10^{-5}$, and $5.34\times 10^{-4}$ for the four cases, which reveal the emergence of strong phase correlations between the two oscillators. In Fig.~\ref{fig1}(e), we show the $10^4$ random results of $\theta_1$ and $\theta_-$
for the case i. The horizontal axis represent the $i$-th random result ($i=1,2,3,...,10^{4}$).
We can determine that synchronization in the CMM system is robust to thermal noises with the protection of the phase correlation.
In addition, we emphasize that the mixture of two (or more) limit cycles never emerges in our simulation results (the considered parameters are not limited to the four points shown), which means that the adjacent attractors are too distant in the phase space. 
Therefore, the noises can not support the mechanical modes jumping from one stable solution to another.

These results prove that a strong phase correlation is established between two vibrational modes, leading to the synchronization between them very robust against thermal noises. It thus suggests that synchronization can be well studied in the noiseless case, which is considered in the following sections.

%%%%%%%%%%%%%%%%%%%%%%%%%%%%%%%%%%%%%%%%%%%%%%%%%%%%%%%%%%%%%%%%%%%%%%%%%%%%%%%%%5

\section{Synchronization phase transition}

For the system under study, it does not exhibit chaotic behavior, which occurs only under an extremely strong driving field. Therefore, we can characterize synchronization in terms of the phase difference~\cite{Weiss013043}: $\mathcal{P}(t)=\cos[\theta_{1}(t)-\theta_{2}(t)]$, where $\theta_{j}$ is the phase of the $j$-th mechanical oscillator, and $\mathcal{P}=-1$, $0$ and $1$ correspond to the $\pi$-phase, non- and zero-phase synchronization, respectively.

\begin{figure}[tp]
{\subfigure{\includegraphics[width=4.2cm,height=4.3cm]{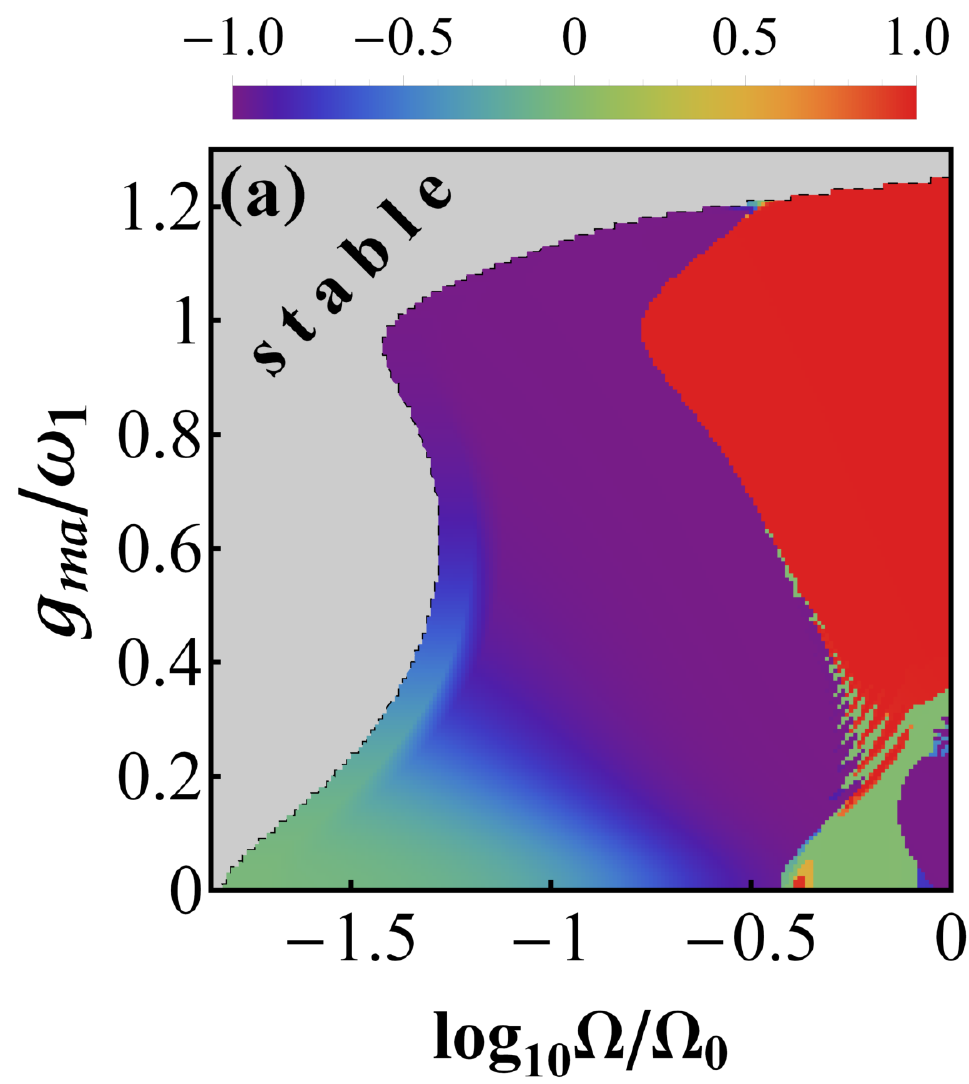}}
\subfigure{\includegraphics[width=4.2cm,height=4.3cm]{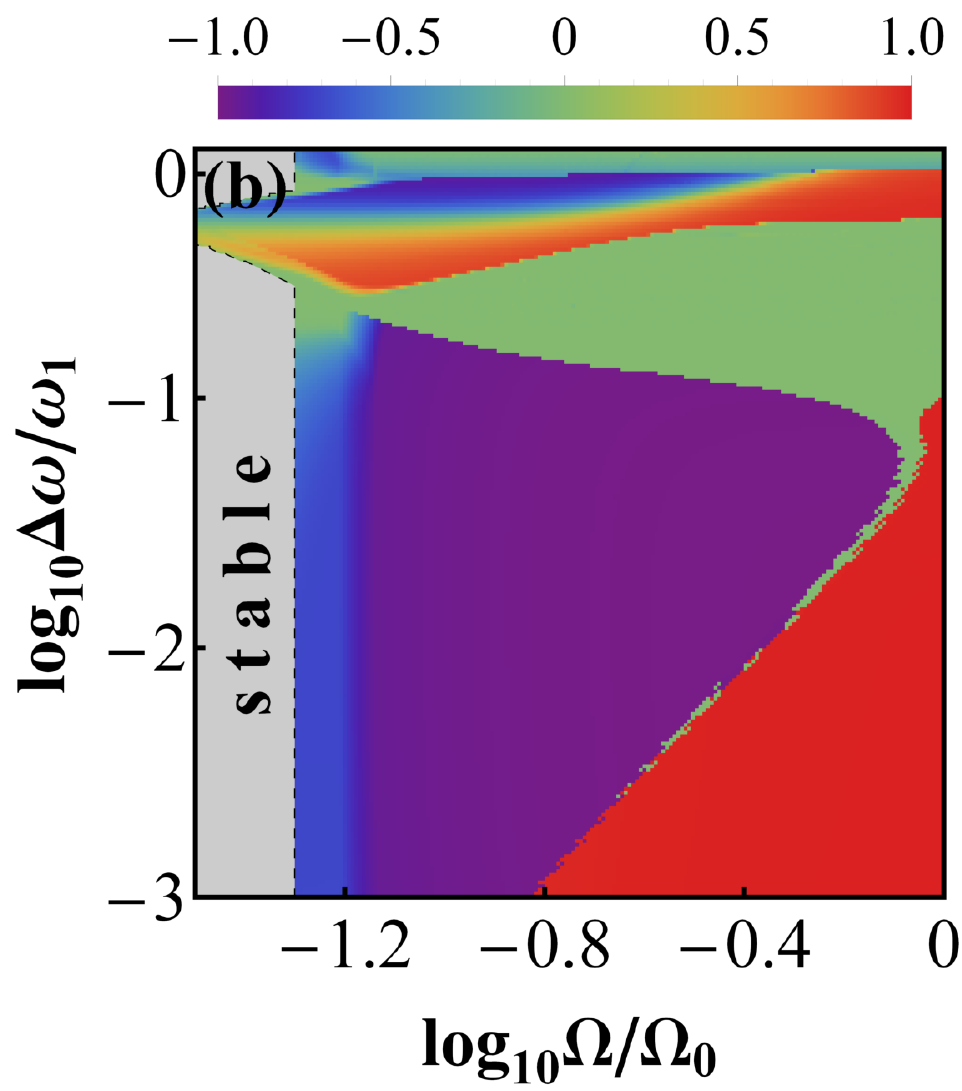}}}
\caption{Synchronization phase diagram as a function of (a) Rabi frequency $\Omega$ and coupling $g_{ma}$; (b) Rabi frequency $\Omega$ and mechanical frequency difference $\Delta\omega=\omega_{2}-\omega_{1}$. We take $\Delta\omega=0.01\omega_{1}$ in (a), $g_{ma}=0.5\omega_{1}$ in (b), and $\Delta_{a}=\Delta_{m}\simeq-\omega_{1}$ in both plots. The gray areas denote the stable regime when the QLEs only have asymptotic steady state solutions, and they can be determined by the Lyapunov stability criterion. The other parameters are the same as in Fig.~\ref{fig1}.}
\label{f2}
\end{figure}

The synchronization phase diagram is shown in Fig.~\ref{f2} by taking time average of $\mathcal{P}(t)$ for a sufficiently long time interval ensuring stable values~\cite{explain2}, i.e., $t\in[9/\gamma_{1},19/\gamma_{1}]$.
Clearly, the $\pi$- and zero-phase synchronizations are present in a large parameter regime, and a prominent phase transition of synchronization appears for a sufficiently strong (small) coupling $g_{ma}$ (mechanical frequency difference $\Delta\omega$).
The synchronization is quite robust and even close to the phase-transition boundary, thermal noises and random initial conditions have negligible impact on the synchronization.
Our theoretical analysis indicates that the system is bistable or even multistable. However, Fig.~\ref{f2} displays only one of the steady states of the limit cycles. This is because we do not traverse all possible initial states, but only assume the system is in a thermal initial state (see Sec. V.).
It is worth noting that, the cavity-magnon coupling $g_{ma}$ (a controllable parameter that can be tuned in a wide range) can effectively modulate the synchronization phase, c.f. Fig.~\ref{f2}(a).  For a moderate value of $g_{ma}$, the two mechanical modes can be $\pi$-phase synchronized even for a large frequency difference $\Delta\omega>0.1\omega_{1}$ and at a low driving power of $83$~$\mu$W, as shown in Fig.~\ref{f2}(b).  This reveals a distinct advantage of the CMM system for realizing and modulating synchronization compared with other systems.

\section{Mechanism of synchronization and multistable synchronized limit cycles}

In order to explain the complex limit cycle dynamics of the system, we take the slowly varying amplitude (SVA) equations approach~\cite{Li013802,Marquardt103901}, and study the long-time dynamics of the two mechanical oscillators in the frame rotating at a fast reference frequency $\bar{\omega}$, i.e.,
$b_{j}(t)=\beta_{j}^{s}+B_{j}e^{-i\bar{\omega} t}$ (see appendix A), where $\beta_{j}^{s}$ are the equilibrium positions, $B_{j}$ are slowly varying complex amplitudes, and $\bar{\omega}=(\omega_{1}+\omega_{2})/2$. Substituting it into the noiseless Langevin equations, we obtain the formal solutions of the cavity and magnon modes, which can be expressed as the sum of a series of sidebands at the frequencies of $n \bar{\omega}$, with $n$ being an integer. Substituting these solutions into the equations of the oscillators, we obtain the following amplitude equations (see appendix A for the detailed derivation):
\begin{eqnarray}
\dot{B}_{j}=-\left[i(\omega_{j}-\bar{\omega})+\gamma_{j}\right]B_{j}
-i\frac{g_{j}F}{\tilde{g}}(g_{1}B_{1}+g_{2}B_{2}), \,\,\,  \label{AEs1} 
%-iFg_{j}\tilde{g}^{-1}(g_{1}B_{1}+g_{2}B_{2}), \label{AEs}
\end{eqnarray}
where $\tilde{g}=\sqrt{g_{1}^{2}+g_{2}^{2}}$, and the dimensionless function
$F(\Delta_{a,m},\kappa,g_{1,2,ma},\vert B\vert,\Omega)=\tilde{g}|\tilde{B}|^{-1}\sum_{n}M_{n}M_{n+1}^{*}$, with $\tilde{B}=\sum_{j}g_{j}B_{j}$.
$M_{n}$ is the amplitude of the $n$-th mechanical sideband, which can be determined via iterative computation or other numerical methods.
Equation~\eqref{AEs1} indicates that the backaction of the cavity-magnon system on the dynamics of the mechanical oscillators is fully manifested in the $F$-function, which renormalizes the frequencies and dissipations, and more importantly, provides an effective coupling between the two oscillators.

By rewriting Eq.~\eqref{AEs1} in terms of the modulus $I_{j}$ and phase $\theta_{j}$ of the complex amplitude $B_{j}=I_{j}e^{i\theta_{j}}$, we obtain the following Kuramoto-like equations (KLEs) \cite{Acebron137}:
\begin{eqnarray}
\dot{I}_{j} &=&\Gamma_j I_{j}
+\frac{g_{1}g_{2}}{\tilde{g}}\left(F_{i}\cos\theta_{-}+\dfrac{F_{r}}{(-1)^j}\sin\theta_{-}\right)I_{3-j}, \notag \\
\dot{\theta}_{-} &=& \frac{g_{1}g_{2}}{\tilde{g}}
\left(F_{r}\cos\theta_{-}\frac{I_{1}^{2}-I_{2}^{2}}{I_{1}I_{2}}
-F_{i}\sin\theta_{-}\frac{I_{1}^{2}+I_{2}^{2}}{I_{1}I_{2}}\right) \notag \\
&&+\frac{g_{2}^{2}-g_{1}^{2}}{\tilde{g}}F_{r}+\Delta\omega,
\label{KLE}
\end{eqnarray}
where $\Gamma_j={g_{j}^{2}F_{i}}/{\tilde{g}}-\gamma_{j}$, $F=F_{r}+iF_{i}$ and the phase difference $\theta_{-}=\theta_{1}-\theta_{2}$. The KLEs provide us a powerful tool to describe the self-sustained mechanical oscillations. %whose synchronization dynamics we explore here. 
They can be further simplified to stationary equations by setting the derivatives to zero, which describe two synchronized oscillators as two amplitude-stable limit cycles will be of a constant phase difference. 
The stationary $F$-function $F^{s}$ can be written as a function of the stationary modulus $I_{j}^{s}$, i.e.,
\begin{eqnarray}
F_{r}^{s} &=& \frac{\tilde{g}\left((g_{1}^{2}\gamma_{2}-g_{2}^{2}\gamma_{1})R^{s}+g_{1}g_{2}(\gamma_{2}-\gamma_{1}{R^{s}}^{2})\cos\theta_{-}^{s}\right)}
{g_{1}g_{2}(g_{2}^{2}+2g_{1}g_{2}R^{s}\cos\theta_{-}^{s}+g_{1}^{2}{R^{s}}^{2})\sin\theta_{-}^{s}}, \notag \\
F_{i}^{s} &=& \frac{\tilde{g}(\gamma_{2}+\gamma_{1}{R^{s}}^{2})}{g_{2}^{2}+2g_{1}g_{2}R^{s}\cos\theta_{-}^{s}+g_{1}^{2}{R^{s}}^{2}},
\label{AEs}
\end{eqnarray}
where we define $R^{s}=I_{1}^{s}/I_{2}^{s}$ for convenience.
Substituting Eqs.~\eqref{AEs} into Eq.~\eqref{KLE} yields the following state constraint equation on $R^{s}$ and $\theta_{-}^{s}$:
\begin{eqnarray}
\Delta\omega\sin\theta_{-}^{s}+(\gamma_{1}+\gamma_{2})\cos\theta_{-}^{s}=\frac{g_{2}\gamma_{1}}{g_{1}}R^{s}+\frac{g_{1}\gamma_{2}}{g_{2}R^{s}},
\label{DBC}
\end{eqnarray}
which determines the behavior of the stationary synchronization of the two oscillators.
Note that the above constraint equation depends only on the mechanical system but not on the cavity-magnon system \cite{explain1}. Hence, the constraint of the synchronization is essentially an intrinsic property of the two oscillators. The solutions of Eq.~\eqref{DBC}, manifested as the identical red lines in Fig.~\ref{f3}(a)-(f), are thus the \textit{necessary conditions} for the synchronization, which are satisfied by all allowed synchronization states under the given parameters, while any other states outside the red lines are actually the unstable states of the limit cycles.   In particular, the perfect zero-phase synchronization $\theta_{-}^{s}=0$ requires $R^{s}=\frac{g_{1}\gamma_{2}}{g_{2}\gamma_{1}}$. By contrast, the perfect $\pi$-phase synchronization is unattainable for the conventional parameters, as it requires $R^{s}=-\frac{g_{1}\gamma_{2}}{g_{2}\gamma_{1}}$.
For a given $\theta_{-}^{s}$, the solution of $R^{s}$ is symmetric, and $\theta_{-}^{s}$ has a single maximum at $R^{s}=\frac{g_{1}}{g_{2}}\sqrt{\frac{\gamma_{2}}{\gamma_{1}}}$, which yields an optimal $\mathcal{P}$ for the $\pi$-phase synchronization, i.e.,
$\mathcal{P}_{\pi}^{\rm opt}=\frac{2(\gamma_{1}+\gamma_{2})\sqrt{\gamma_{1}\gamma_{2}}-\Delta\omega\sqrt{\Delta\omega^{2}+(\gamma_{1}-\gamma_{2})^{2}}}{\Delta\omega^{2}+(\gamma_{1}+\gamma_{2})^{2}}$.
Apparently, $\Delta\omega\gg\gamma_{1,2}$ is the basic condition for the occurrence of the $\pi$-phase synchronization.

\begin{figure}[]
\centering
\includegraphics[width=8.5cm]{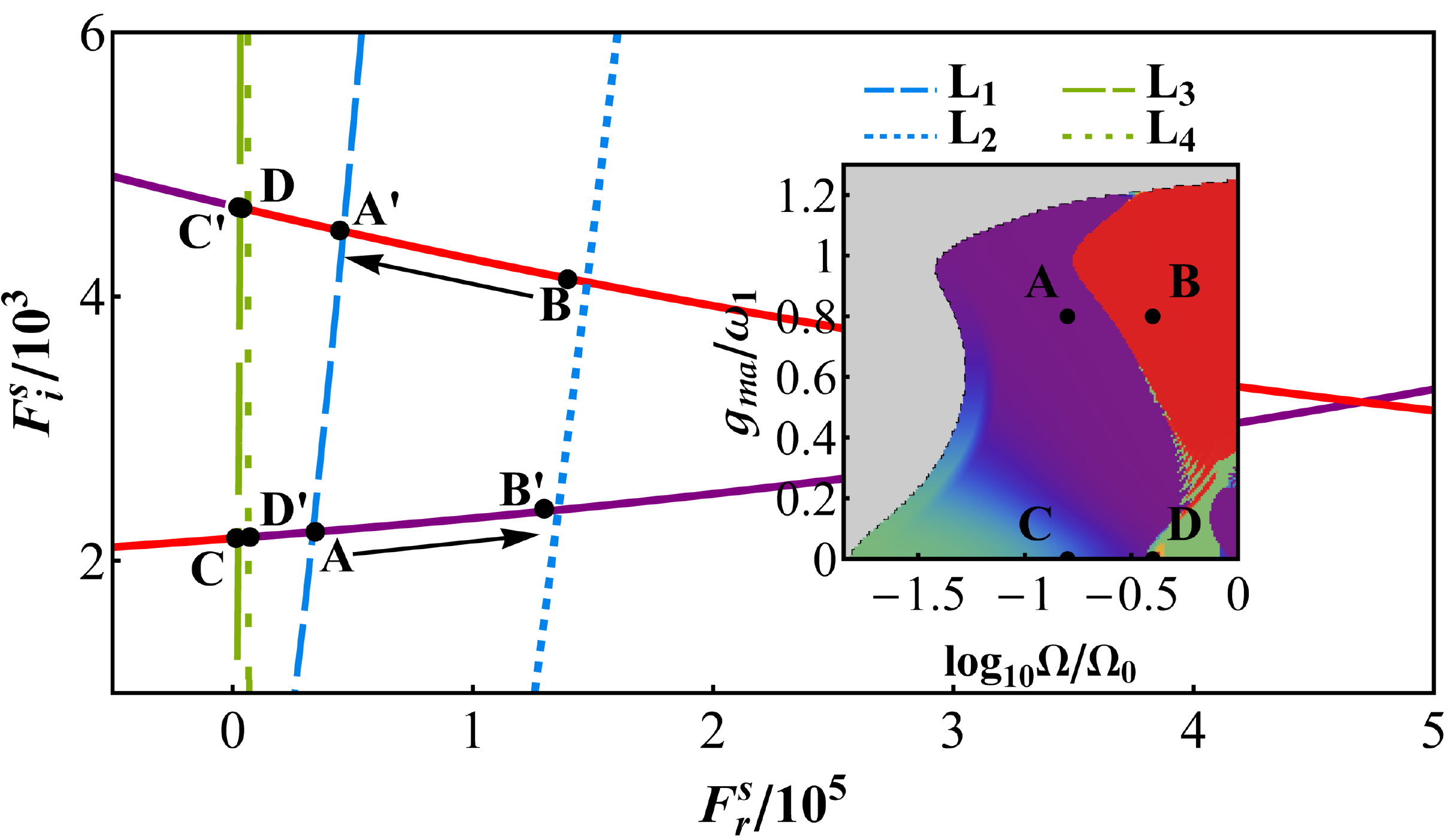}
\caption{Parametric plot of $F$. The red (purple) lines denote the steady-state range of the zero-phase ($\pi$-phase) synchronization in the coordinates of $F_{r}^{s}$ and $F_{i}^{s}$. The lines ($L_{1}$-$L_{4}$) represent the dynamic range of $F$, with $g_{ma}/\omega_{1}=0.8$ and $\Omega/\Omega_{0}=10^{-0.8}$ for $L_{1}$; $g_{ma}/\omega_{1}=0.8$ and $\Omega/\Omega_{0}=10^{-0.4}$ for $L_{2}$; $g_{ma}/\omega_{1}=0$ and $\Omega/\Omega_{0}=10^{-0.8}$ for $L_{3}$; and $g_{ma}/\omega_{1}=0$ and $\Omega/\Omega_{0}=10^{-0.4}$ for $L_{4}$. Inset shows the points $A$, $B$, $C$ and $D$ marked in the synchronization phase diagram, of which the parameters are the same as $L_{1}$, $L_{2}$, $L_{3}$ and $L_{4}$, respectively.
\label{fig2}}
\end{figure}

\begin{figure}[t]
\includegraphics[height=5.8cm]{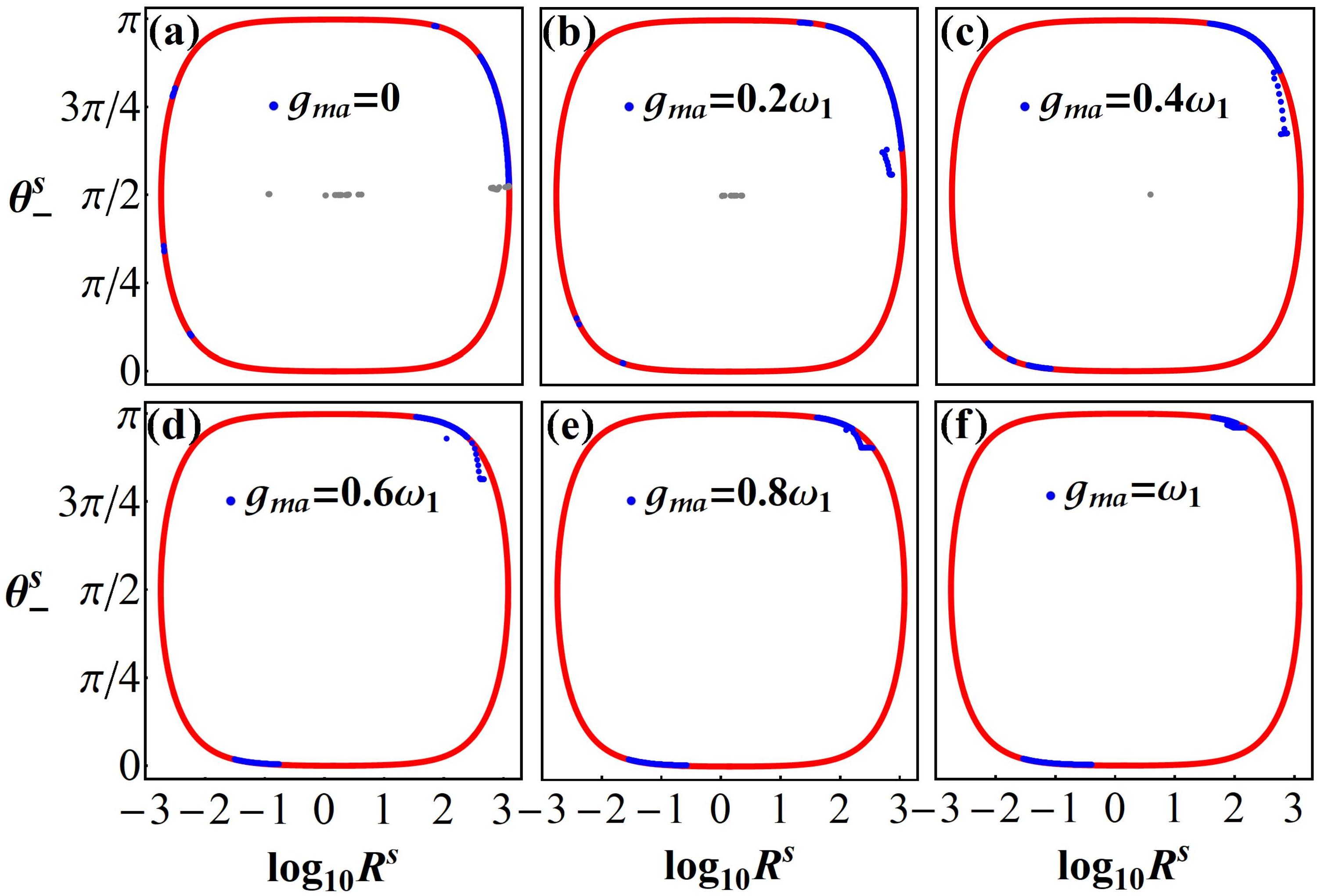}
\caption{Steady-state distribution of $R^{s}$ and $\theta_{-}^{s}$ with different values of $g_{ma}$. 
The red line is the solution of Eq.~(\ref{DBC}), and each blue point corresponds to a pixel in Fig. \ref{f2}(a).
The gray points represent the non-synchronization states, of which the phase difference $\theta_-$ are nonstationary, but their time average $\theta_-^s$ are around $\pi/2$. The parameters are the same as in Fig.~\ref{f2}(a). }
\label{f3}
\end{figure}

Utilizing the analytical expression of Eq. (\ref{AEs}), we now discuss in detail the mechanism of synchronization phase transition.
We still study the synchronization dynamics in the long-time limit.
As shown in Fig.~\ref{fig2}, the steady-state range $F^{s}$ (red and purple lines obtained by solving Eqs.~(\ref{AEs}-\ref{DBC})) and the dynamic range of $F$ (lines $L_{1}$-$L_{4}$ obtained by solving Eq.~(\ref{mabs})) is plotted.
The intersection point of $F^{s}$ and $F$ indicates that the limit cycles have a steady-state solution.
The points $A$, $B$, $C$ and $D$ (also marked in the inset) represent the value of the steady-state solution $F^{s}$ at the corresponding parameter of the phase diagram.
These points are obtained by first simulating the noiseless Langevin equations with the thermal initial states
for a sufficiently long time ensuring stable values (i.e., $\theta_{-}^{s}$ and $R^{s}$), then substituting them into the $F^{s}(\theta_{-}^{s},R^{s})$ function.
Clearly, our theoretical approach fits perfectly with the numerical simulation, and more importantly, it reveals the bistability of the limit cycles, i.e., the points $A'$-$D'$.
Here, the points $A'$-$D'$ are obtained by simulating the noiseless Langevin equations under appropriate initial conditions.
To be specific, starting with the final state of $A$ and using the parameters of $B$ (or increasing the drive power), we can obtain a new phase $B'$. 
Similarly, starting with the final state of $B$ and using the parameters of $A$ (or lowering the drive power), we achieve a new phase $A'$.
Therefore, the phase transition is essentially induced by the initial thermal states, which will be different if under different (appropriate) initial conditions.

\section{Modulation of synchronization}

\begin{figure}[t]
{~~~~~~~\subfigure{\includegraphics[width=6.2cm]{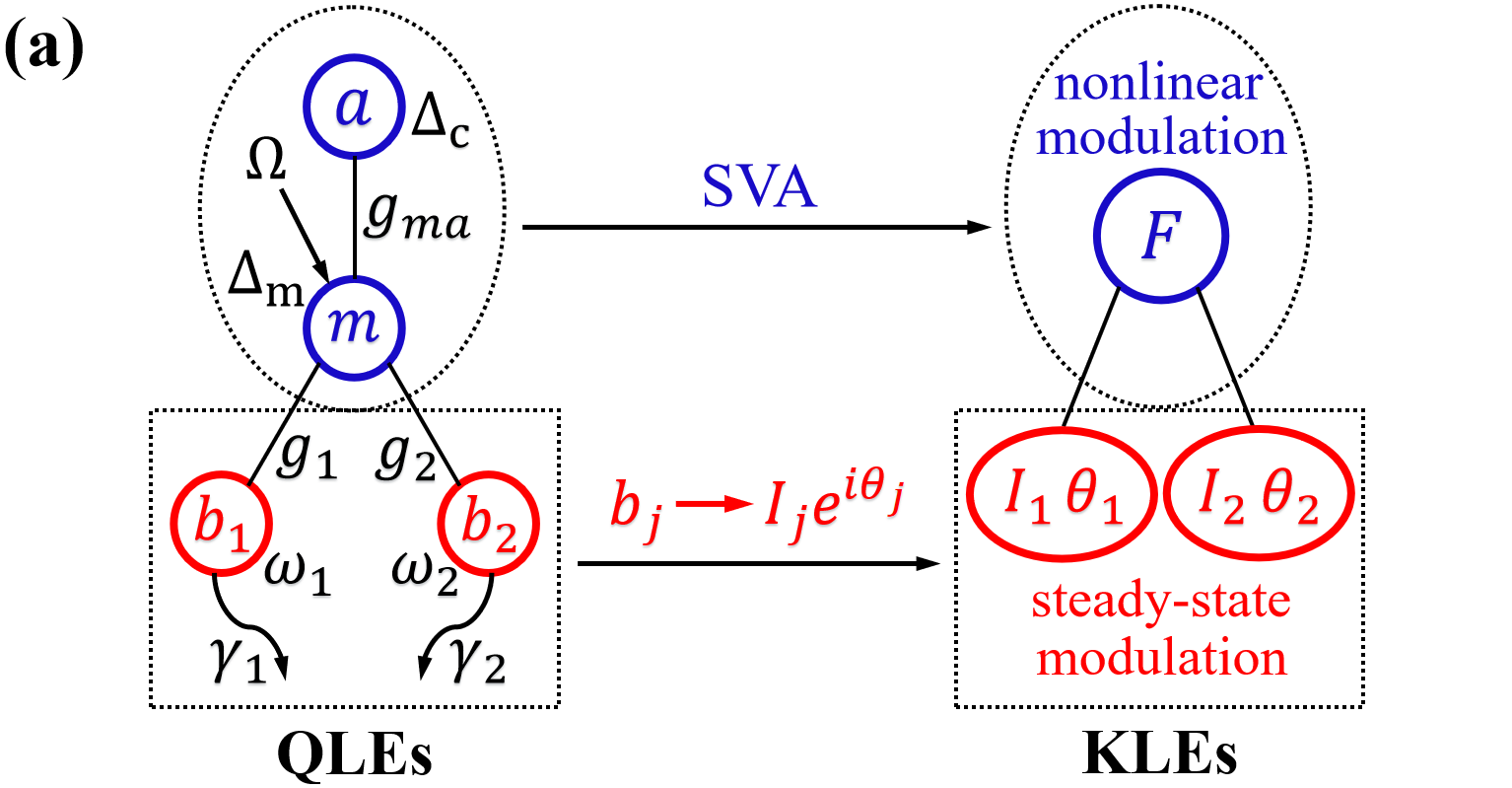}}\vspace{-0.5ex}\\
\subfigure{\includegraphics[width=7cm]{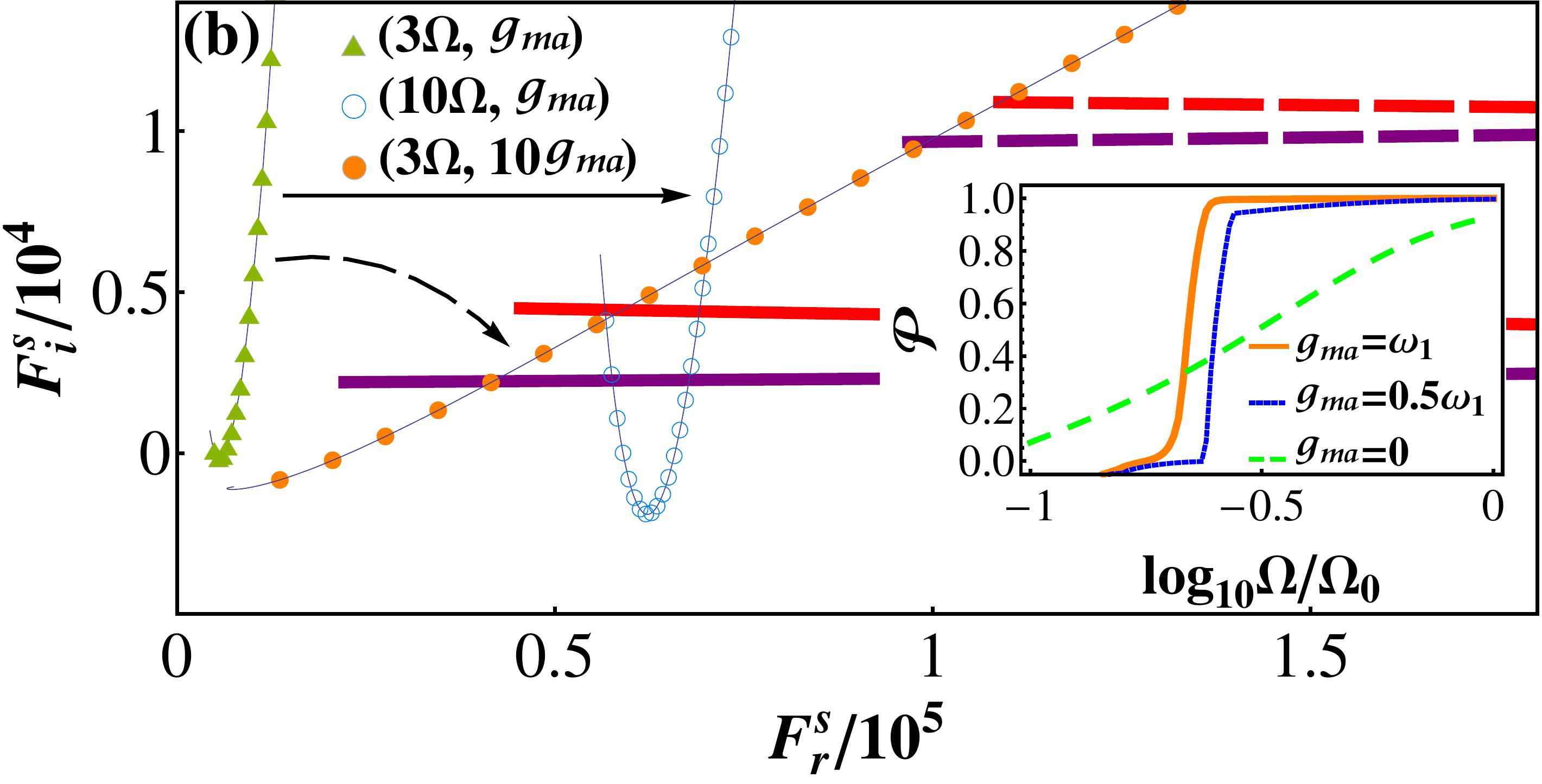}}\vspace{-0.5ex}\\
~\subfigure{\includegraphics[width=7cm]{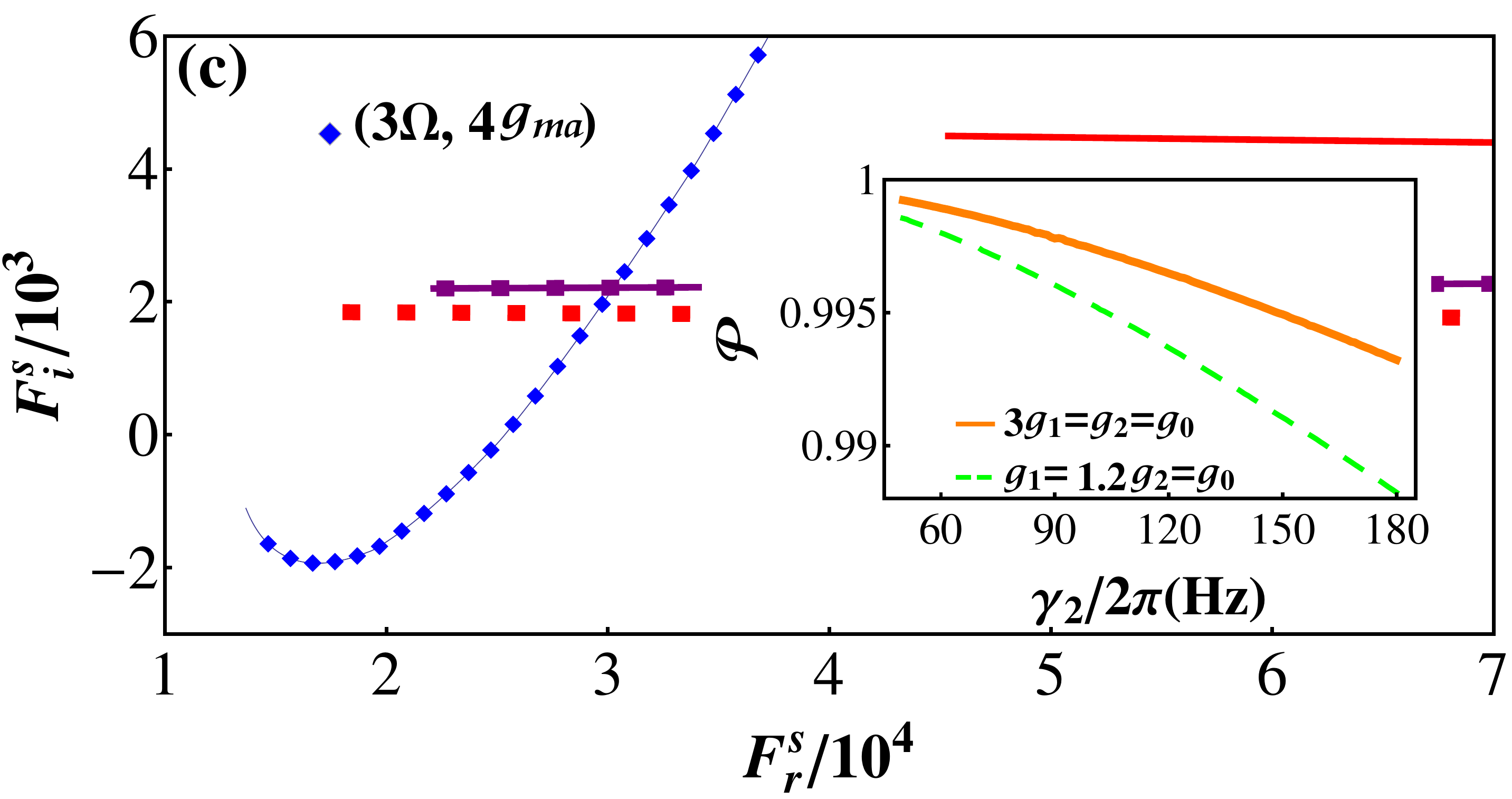}}}
\caption{(a) Schematic diagram of synchronization modulation. 
(b) Nonlinear modulation by controlling the cavity-magnon system  (i.e., via controlling $F$).
The red (purple) lines denote the zero-phase ($\pi$-phase) synchronization regime with $g_{1}/2\pi=60$~mHz, $g_{2}/2\pi=50$~mHz (solid lines), and $g_{1}/2\pi=16$~mHz, $g_{2}/2\pi=18$~mHz (dashed lines and inset). Inset shows the $\mathcal{P}-\Omega$ relation for different values of $g_{ma}$. (c) Steady-state modulation by controlling the mechanical system (i.e., via controlling $F^{s}$). 
The red (purple) lines denote the zero-phase ($\pi$-phase) synchronization regime with $\gamma_{1}/2\pi=100$~Hz, $\gamma_{2}/2\pi=150$~Hz (solid lines), and
 $\gamma_{1}/2\pi=100$~Hz, $\gamma_{2}/2\pi=60$~Hz (dotted lines).
Inset shows the impact of $\gamma_{2}$ on the synchronization, where $g_{0}/2\pi=60$~mHz.
In (b) and (c), we take $\Omega=0.1\Omega_{0}$ and $g_{ma}=0.1\omega_{1}$.
The other parameters are the same as in Fig.~\ref{f2}.}
\label{f4}
\end{figure}

According to the value of $g_{ma}$, we classify the states of the mechanical oscillators (the asymptotic steady states marked in gray are excluded) in the phase diagram of Fig.~\ref{f2}(a) and plot their characteristic variables $\theta_{-}^{s}$ and $R^{s}$ in Fig.~\ref{f3}(a)-(f). The blue scatter points are ``hitched'' by the solution of the constraint equation, as excepted, but they do not completely smear the red lines, confirming that the constraint equation is only a necessary condition. As $g_{ma}$ increases, the blue points tend to distribute to both poles, implying that the system has a distinct feature of zero- or $\pi$-phase synchronization. This indicates that the synchronization can be enhanced and modulated by adjusting the cavity-magnon coupling rate. The synchronization properties beyond the constraint equation are reflected in the aforementioned $F$-function, which is entirely determined by the cavity-magnon system. This suggests dividing the whole system into two parts, as sketched in Fig.~\ref{f4}(a): the mechanical system that constrains the range of the legal synchronization states, corresponding to the {\it steady-state modulation}; and the cavity-magnon system that selects which kind of the synchronization state that can be finally obtained, corresponding to the {\it nonlinear modulation}. These two types of modulation are mutually independent, which greatly simplifies the procedures for synchronizing the oscillators to a given target phase. Specifically, the procedures are summarized as follows: i) Solving $R^{s}$ from the constraint equation Eq.~\eqref{DBC} with a given $\theta^{s}_-$; ii) Substituting $\theta^{s}_-$ and $R^{s}$ into Eqs.~\eqref{AEs} to obtain the conditions that $F$ should fulfil, denoted as $F^{s}$; iii) Adjusting relevant parameters of the cavity-magnon system to satisfy $F=F^{s}$, corresponding to the KLEs having steady-state solutions and thus the occurrence of the target synchronization state.

Due to the nonlinearity of the $F$-function, the last procedure is more easily realized by checking the intersection points after plotting $F$ and $F^{s}$ in the parametric space, as shown in Fig.~\ref{f4}(b).   As we are interested in prominent synchronization phenomena, the phase difference is restricted to a small range satisfying $|\mathcal{P}|>0.995$ and the nonlinear modulation is realized by controlling $F$ via changing $\Omega$ and $g_{ma}$. When the driving power and the coupling rate are small (green triangles), there will be no solutions. 
As the power increases, the curve is shifted along the $F_{r}^{s}$ axis, leading to the appearance of multistable synchronized limit cycles (blue circles).
Comparing with the blue circles (with Rabi frequency $\Omega_{0}$ and cavity-magnon coupling rate $0.1\omega_{1}$), the orange dots (with Rabi frequency $0.3\Omega_{0}$ and cavity-magnon coupling rate $\omega_{1}$) can produce both the zero- and $\pi$-phase synchronizations even for relatively small couplings, e.g., $g_{1}/2\pi=16$~mHz and $g_{2}/2\pi=18$~mHz (dashed lines), which can be easily achieved in the CMM experiments \cite{Tang,Jie18,Davis,Jie22}. This benefits from a new mechanism of synchronization: as $g_{ma}$ increases, the curve is rotated around the original point, which can sweep over a much wider area in the parametric space. Inset shows how the coupling $g_{ma}$ modulates the synchronization. Clearly, increasing the cavity-magnon coupling can significantly enhance the synchronization. In Fig.~\ref{f4}(c), we explore the steady-state modulation via controlling $F^{s}$ realized by altering $\gamma_{j}$. The results indicate that the zero-phase synchronization can be enhanced by reducing the dissipation rates, as more clearly shown in the inset.

\section{Conclusions}

We present a new mechanism of synchronizing mechanical oscillators in a CMM system exploiting the nonlinear magnetostriction. We find that a strong phase correlation can be established between two mechanical oscillators, leading to their synchronization which is robust against thermal noise. We also develop a theoretical framework to analyze the synchronization and determine the active role the cavity-magnon coupling plays in enhancing and modulating the synchronization. All of these indicate that the highly controllable and tunable CMM system can be a promising new platform for studying and modulating synchronization. The work can be extended straightforwardly to study synchronization between two or multi YIG spheres. It can also be applied to other systems that share a similar Hamiltonian as the CMM system, e.g., synchronizing two mechanical oscillators in exciton-optomechanics systems~\cite{eom1,eom2,eom3}. Synchronized mechanical oscillators can be exploited to achieve the synchronization between two optical cavities, e.g., by means of an opto-magnomechanical configuration~\cite{QST,Fan033507}, and between two atomic ensembles by further coupling each cavity to an atomic ensemble~\cite{Fan023501}. This provides possibility to distribute synchronization or quantum states in a complex quantum network~\cite{Li022204}.

%%%%%%%%%%%%%%%%%%%%%%%%%%%%%%%%%%%%%%%%%%%%%%%%%%%%%%%%%%%%%%%%%%%%%%%%%%%%%%%%%%%%%%%%%%%%%%%%%%%%%%%%%%%%%%%%%%%%%%%%%

\begin{acknowledgments}

This work has been supported by National Key Research and Development Program of China (Grant No. 2022YFA1405200) and National Natural Science Foundation of China (Grant Nos. 92265202, 11704205, 12304389 and 12074206). We also acknowledge the support of the European Union Horizon 2020 Programme for Research and Innovation through Project No. 732894 (FET Proactive HOT) and the Scientific Research Foundation of NEU (Grant No. 01270021920501*115).

\end{acknowledgments}
%%%%%%%%%%%%%%%%%%%%%%%%%%%%%%%%%%%%%%%%%%%%%%%%%%%%%%%%%%%%%%%%%%%%%%%%%%%%%%%%%%%%%%%%%%%%%%%%%%%%%%%%%%%%%%%%%%%%%%%%%

\appendix

\section{Derivation of the slowly varying amplitude equations}

The noiseless Langevin equations of the system are given by
\begin{eqnarray}
\dot{a} &=& -(i\Delta_a+\kappa_a)a-ig_{ma}m, \notag \\
\dot{m} &=& -i(\Delta_m+\kappa_m)m-ig_{ma}a  \notag \\
&&-\sum_{j=1,2} ig_jm(b_j^*+b_j)+\Omega, \label{CLE2} \\ 
\dot{b}_j &=& -(i\omega_j+\gamma_j)b_j-ig_j\vert m\vert^2. \notag
\end{eqnarray}
We now consider the self-sustaining solution of the mechanical modes.
After an initial transient regime, the dynamics of the mechanical modes have the following form \cite{Marquardt103901,Rodrigues2010}:
\begin{equation}
b_{j}(t)=\beta_{j}^{s}+B_{j}e^{-i\bar{\omega} t},
\label{ss solution}
\end{equation}
where $\beta_{j}^{s}$ are the equilibrium positions, $B_{j}$ are slowly varying complex amplitudes, and the reference frequency $\bar{\omega}=(\omega_{1}+\omega_{2})/2$.
Here, the chaotic motion of the mechanical modes are neglected, as it occurs only at extremely large driving powers. %which is not physically meaningful for the CMM we considered.
Substituting the solution Eq.~(\ref{ss solution}) into Eq.~(\ref{CLE2}), we have 
\begin{eqnarray}
\dot{m} &=& -i(\Delta_{m}+\tilde{\beta}^{s}+2|\tilde{B}|\cos(\bar{\omega}t-\varphi))m \notag \\
&&-\kappa_{m}m-ig_{ma}a+\Omega,
\label{mEq}
\end{eqnarray}
where $\tilde{\beta}^{s}=\sum_{j}g_{j}({\beta_j^{s}}^{*}+\beta_j^{s})$ and $\tilde{B}=\sum_{j}g_{j}B_{j}=|\tilde{B}|e^{i\varphi}$.
Eq.~(\ref{mEq}) has the formal solution:
\begin{eqnarray}
m(t) &=& \int_{0}^{t}d\tau e^{\int_{\tau}^{t}d\tau'[-i(\Delta_{m}+\tilde{\beta}^{s}+2|\tilde{B}|\cos(\bar{\omega}\tau'-\varphi))-\kappa_{m}]} \notag \\
&&\times(-ig_{ma}a(\tau)+\Omega).
\label{mFSol}
\end{eqnarray}
Note that the order of the characteristic time corresponding to the dynamics of the amplitude $|\tilde{B}|$ is $\gamma_{j}$, which is much slower than the fast oscillations at $\bar{\omega}$, and one thus can treat $|\tilde{B}|$ as a constant in the integral over $\tau'$ in Eq.~(\ref{mFSol}). We then have
\begin{eqnarray}
m(t) &=& e^{-i\frac{2|\tilde{B}|}{\bar{\omega}}\sin(\bar{\omega}t-\varphi)}\sum_{n}J_{n}(\frac{2|\tilde{B}|}{\bar{\omega}}) \notag \\
&&\times\int_{0}^{t}d\tau e^{in(\bar{\omega}\tau-\varphi)-[i(\Delta_{m}+\tilde{\beta}^{s})+\kappa_{m}](t-\tau)} \notag \\
&&\times(-ig_{ma}a(\tau)+\Omega),
\label{mFSol2}
\end{eqnarray}
where we use the Jacobi-Anger expansion $e^{i\frac{2|\tilde{B}|}{\bar{\omega}}\sin(\bar{\omega}\tau-\varphi)}=\sum_{n}J_{n}(\frac{2|\tilde{B}|}{\bar{\omega}})e^{in(\bar{\omega}\tau-\varphi)}$, and $J_{n}$ is the $n$-th Bessel function of the first kind.
The nonlinear interaction will lead to the magnon mode exhibiting complex dynamics accompanied by higher-order sidebands, satisfying the following form:
\begin{equation}
m(t)=\sum_{n}M_{n}e^{in(\bar{\omega}t-\varphi)}.
\label{mSol}
\end{equation}
Substituting Eq.~(\ref{mSol}) into Eq.~(\ref{CLE2}), we have the solution:
\begin{equation}
a(t)=\sum_{n}\frac{-ig_{ma}M_{n}}{i(\Delta_{a}+n\bar{\omega})+\kappa_{a}}e^{in(\bar{\omega}t-\varphi)}.
\label{aFSol}
\end{equation}
Inserting Eq.~(\ref{aFSol}) into Eq.~(\ref{mFSol2}) and comparing the coefficients on both sides, we can solve the corresponding equation by the iterative method.
We finally determine the following iterative equation
\begin{equation}
M_{n}=\sum_{k,l}\frac{J_{n-k-l}(-\frac{2|\tilde{B}|}{\bar{\omega}})J_{k}(\frac{2|\tilde{B}|}{\bar{\omega}})(\Omega\delta_{l,0}-\frac{g_{ma}^{2}M_{l}}{i(\Delta_{a}+l\bar{\omega})+\kappa_{a}})}{i[\Delta_{m}+\tilde{\beta}^{s}+(k+l)\bar{\omega}]+\kappa_{m}}.
\label{Mn}
\end{equation}
$M_{n}$ can be finally determined after the errors converge to an acceptable range through multiple iterations, i.e., $|M_{n}^{j+1}-M_{n}^{j}|<\epsilon$.
The magnon excitation term can be written as
\begin{equation}
|m(t)|^{2}=\sum_{nn'}M_{n+n'}M_{n'}^{*}e^{in(\bar{\omega}t-\varphi)}.
\label{mabs}
\end{equation}
Substituting Eq.~(\ref{mabs}) into Eq.~(\ref{CLE2}), we obtain
\begin{eqnarray}
\beta_{j}^{s}&=&\frac{-ig_{j}}{i\omega_{j}+\gamma_{j}}\sum_{n}|M_{n}|^{2}, \label{zero} \\
\dot{B}_{j}&=&-[i(\omega_{j}-\bar{\omega})+\gamma_{j}]B_{j}-ig_{j}\tilde{g}^{-1}(g_{1}B_{1}+g_{2}B_{2})F, \notag
\end{eqnarray}
where $\tilde{g}=\sqrt{g_{1}^{2}+g_{2}^{2}}$, and the dimensionless auxiliary function $F$ is defined as
\begin{equation}
\begin{split}
F=\frac{\tilde{g}}{|\tilde{B}|}\sum_{n}M_{n}M_{n+1}^{*}.
\end{split}
\label{F}
\end{equation}

\section{Synchronization of two YIG spheres in cavity magnomechanics}

Here we study on the synchronization of two mechanical vibrational modes of two YIG spheres that are spatially separated, e.g., placed at the antinodes of the magnetic field of the same cavity mode. The realization of such remote synchronization between two or multi YIG spheres would be more attractive, but also more difficult.

The system consists of two YIG spheres, each supporting a magnon mode and a mechanical mode, interacting with a common cavity mode,  which is driven by a microwave field. The Hamiltonian of the system is given by
\begin{eqnarray}
\hat{H}/\hbar &=& \omega_{c}\hat{a}^{\dag}\hat{a}
+\sum_{j=1,2} \omega_{mj}\hat{m}^{\dag}_{j}\hat{m}_{j}+\omega_{j}\hat{b}_{j}^{\dag}\hat{b}_{j} \notag \\
&&+g_{j}\hat{m}^{\dag}_{j}\hat{m}_{j}(\hat{b}_{j}^{\dag}+\hat{b}_{j})
+g_{ma}(\hat{a}^{\dag}\hat{m}_{j}+\hat{m}^{\dag}_{j}\hat{a}) \notag \\
&&+i\Omega_{a} (\hat{a}^{\dag}e^{-i\omega_{0}t}-\hat{a}e^{i\omega_{0}t}),
\label{Hnew}
\end{eqnarray}
where $\Omega_{a}=\sqrt{\frac{2\kappa_{ex}P_0}{\hbar\omega_{0}}}$ denotes the drive-cavity coupling strength, with $\kappa_{ex}$ being the external decay rate of the cavity through the input port and $P_0$ the drive power.
In the frame rotating at the driving frequency $\omega_{0}$, by adding dissipative and input noise terms, we obtain the following QLEs:
\begin{eqnarray}
\dot{\hat{a}} &=& -(i\Delta_{a}+\kappa_{a})\hat{a}-ig_{ma}(\hat{m}_{1}+\hat{m}_{2}) \notag \\
&&+\Omega_{a}+\sqrt{2\kappa_{in}}\hat{a}_1^{in} +\sqrt{2\kappa_{ex}}\hat{a}_2^{in}, \notag \\
\dot{\hat{m}}_{j} &=& -(i\Delta_{j}+\kappa_{j})\hat{m}_{j}-ig_{ma}\hat{a}-ig_{j}\hat{m}_{j}(\hat{b}_{j}^{\dag}+\hat{b}_{j}) \notag \\
&&+\sqrt{2\kappa_{j}}\hat{m}^{in}_{j},\notag \\
\dot{\hat{b}}_{j} &=& -(i\omega_{j}+\gamma_{j})\hat{b}_{j}-ig_{j}\hat{m}^{\dag}_{j}\hat{m}_{j}+\sqrt{2\gamma_{j}}\hat{b}^{in}_{j},
\label{QLEs2}
\end{eqnarray}
where $\kappa_{a}=\kappa_{in}+\kappa_{ex}$ is the total cavity decay rate, with $\kappa_{in}$ being the intrinsic cavity decay rate, and $\Delta_{j}=\omega_{mj}-\omega_{0}$.
The above QLEs can be well approximated by a set of coupled noiseless Langevin equations \cite{Benlloch133601}, considering the synchronization is very robust against thermal noise:
\begin{eqnarray}
\dot{a} &=& -(i\Delta_{a}+\kappa_{a})a-ig_{ma}(m_{1}+m_{2})+\Omega_{a}, \notag \\
\dot{m}_{j} &=& -(i\Delta_{j}+\kappa_{j})m_{j}-ig_{ma}a-ig_{j}m_{j}(b_{j}^{*}+b_{j}),\notag \\
\dot{b}_{j} &=& -(i\omega_{j}+\gamma_{j})b_{j}-ig_{j}\vert m_{j}\vert^2.
\label{NLLEs}
\end{eqnarray}

\begin{figure}[t]
\centering
\includegraphics[width=7.cm]{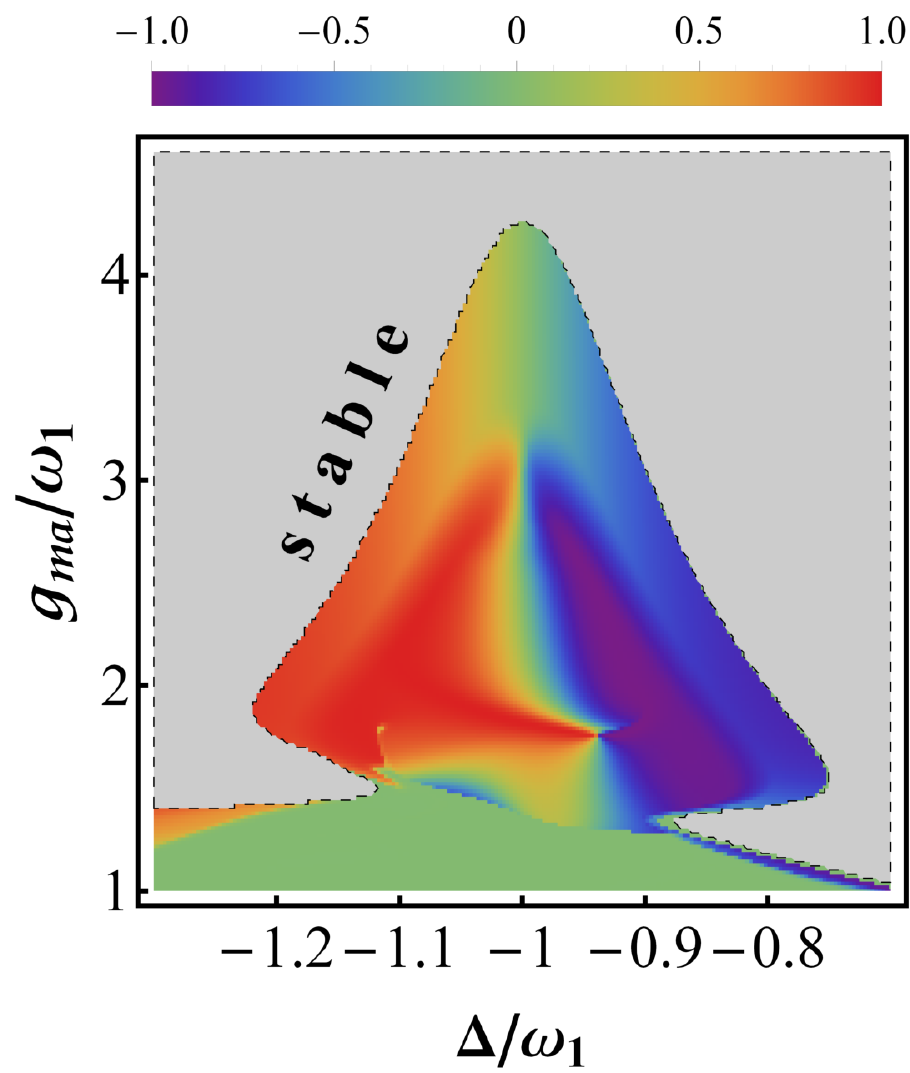}
\caption{Synchronization phase diagram in the detuning-coupling plane. We take $\Delta_{a}=\Delta_{1}=\Delta_{2}=\Delta$, $\kappa_{1}=\kappa_{2}=\kappa_{ex}=2\pi\times1$ MHz, and $P_0=8$ mW. 
The parameters are the same as in Fig.~\ref{f2}(a).
\label{fig3}}
\end{figure}

In Fig. \ref{fig3}, we plot the synchronization phase diagrams by using Eq.~(\ref{NLLEs}) and taking time average of the phase difference $\mathcal{P}(t)=\cos[\theta_{1}(t)-\theta_{2}(t)]$ for a sufficiently long time interval ensuring stable values, i.e., $t\in[9/\gamma_{1},19/\gamma_{1}]$.
In this system of two YIG spheres, the effective coupling between the two mechanical modes is much more indirect (via the mediation of two magnon modes and a common cavity), compared with the case studied in the main text. This makes it more difficult to synchronize two mechanical modes of two YIG spheres, reflected by the fact that the parameter regime for achieving synchronization is much smaller than the single-sphere case.
To achieve zero-phase ($\pi$-phase) synchronization, a much stronger cavity-magnon coupling rate is needed, about $g_{ma}\simeq 2 \omega_{1}$. Such a strong coupling can, however, be easily obtained in cavity magnonic experiments, thanks to the high spin density of YIG. 
This indicates the advantage of the system: the achievable very strong cavity-magnon coupling can effectively enhance and modulate the synchronization of two mechanical modes of either one YIG sphere or two YIG spheres.

%%%%%%%%%%%%%%%%%%%%%%%%%%%%%%%%%%%%%%%%%%%%%%%%%%%%%%%%%%%%%%%%%%%%%%%%%%%%%%%%%%%%%%%%%%%%%


\begin{thebibliography}{99}
\bibitem{explain2} Through numerical simulation, we find that except for the dynamic parameters that are very close to the critical limit (boundary) of asymptotic stability and self-sustaining dynamics, the system will reach a stable state within $\gamma t< 5$. Therefore, the dynamics in the range of $9<\gamma t<19$ can be considered as the long-term stable dynamics of the system.
\bibitem{explain1} The constraint equation depends on the ratio of the magnomechanical couplings $g_1/g_2$, but not on their specific values. Since the two mechanical oscillators are coupled to a common magnon mode, it is reasonable to regard the ratio $g_1/g_2$ as the intrinsic property of the two oscillators.
\end{thebibliography}
\end{document}